\documentclass[%
 reprint,
 nofootinbib,
 amsmath,amssymb,
 aps,
]{revtex4-2}

\usepackage{graphicx}% Include figure files
\usepackage{dcolumn}% Align table columns on decimal point
\usepackage{bm}% bold math
\usepackage [dvipsnames] {xcolor}
\usepackage{hyperref}% add hypertext capabilities
\hypersetup
{
colorlinks=true,
linkcolor=blue,
filecolor=blue,
urlcolor=blue,
citecolor=blue,
}
\usepackage[mathlines]{lineno}% Enable numbering of text and display math
\begin{document}

\preprint{APS/123-QED}

\title{Probing stellar-mass primordial black holes with type Ia supernova microlensing}

\author{Mingqi Sun$^{1}$}
\author{Kai Liao$^{1}$}%
 \email{E-mail:liaokai@whu.edu.cn}

\affiliation{$^{1}$School of Physics and Technology, Wuhan University, Wuhan 430072, China}

\date{\today}% It is always \today, today,

\begin{abstract}
    Gravitationally lensed quasars have served as a powerful tool for studying the composition of dark matter (DM) in lensing galaxies. In this work, we propose a novel method to investigate stellar-mass primordial black holes (PBHs) by using the microlensing effect of strongly lensed Type Ia supernovae (SNe Ia). Using the parameters of the lensed quasar system PG 1115+80, such as convergence, shear, and stellar/dark matter fractions, we generate microlensing magnification maps. A uniform brightness disk model is applied to these maps to evaluate the microlensing amplitude at different stages of the supernova explosion. We extend this analysis by employing the strong lensing parameters derived from the Vera C. Rubin Observatory’s Legacy Survey of Space and Time (LSST) to create extensive image datasets of lensed SNe Ia. Utilizing these datasets and the Kolmogorov-Smirnov (KS) test, we compare two models: (1) the fiducial model, where galaxies are composed of stars and smooth DM, and (2) the alternative model, where galaxies consist of stars and compact DM, specifically PBHs. Our preliminary analysis predicts that at least 60 image datasets are required to distinguish these two scenarios at a 95\% confidence level. Additionally, by incorporating the \textit{Strong Lensing Halo model-based mock catalogs (SL-Hammocks)}, which provide more realistic and precise image data, we refine our prediction to assess the data requirements for distinguishing cases where PBHs constitute fractions x of the total dark matter mass. Our findings indicate that 50, 55, and 65 image datasets, corresponding to compact dark matter fractions of 100\%, 50\%, and 25\% (denoted by x), are necessary to distinguish between the specific models.
\end{abstract}

\maketitle

%\tableofcontents

\section{Introduction}
Dark matter is a hypothetical form of matter that, on a cosmological scale, is approximately five times more abundant than the familiar baryonic matter, such as matter composed of neutrons and protons. On the scale of galaxies, dark matter accounts for about 80\% of a galaxy's total mass. Since Zwicky's studies of galaxy clusters\cite{bertone2018history}, astronomers have widely accepted the gravitational effects of dark matter and applied them extensively in astrophysical research. However, a solid theoretical understanding of the nature of dark matter remains lacking. Current research on dark matter models is primarily divided into two categories. One category involves particle dark matter, such as axions, weakly interacting massive particles (WIMPs), etc\cite{boyarsky2019sterile,oks2021brief}, which are smoothly distributed. The other category considers dark matter in a more compact form, such as massive compact halo objects (MACHOs) and primordial black holes (PBHs)\cite{alcock1992search,bertone2018history}. Research into the nature of dark matter is crucial for the precise construction of cosmological models.

Microlensing is a vital tool for studying the composition of matter in galaxies. Analogous to lensed quasars, strongly lensed supernovae also produce multiple images. On this foundation, microlensing effects are primarily influenced by compact objects, such as stars distributed within the lens galaxies\cite{chang1979flux,gott1981heavy}. Light from a source (e.g., a quasar\cite{vernardos2024microlensing} or supernova\cite{suyu2024strong}) is further split into multiple sub-images by the microlensing effect. These sub-images typically have angular separations on the \(\mu\)-arcseconds scale, which is far below the spatial resolution limit of current telescopes. As a result, we can only infer the specific impact of microlensing by observing anomalies in the strongly lensed images, such as flux variations or time delays.

Current research indicates that typical microlensing signals primarily originate from the distribution of stars within the lens galaxy\cite{schild1990time,falco1991role}. However, studies on the light curves of lensed quasars suggest that incorporating compact objects, such as primordial black holes (PBHs), provides a more comprehensive explanation for the influence of microlensing effect on the light curves\cite{hawkins2020signature,hawkins2020sdss}. A considerable amount of research has utilized galactic microlensing to constrain the abundance of primordial black holes (PBHs) across various mass ranges, such as\cite{blaineau2022new}. Moreover, several previous studies have obtained significant results by using quasar microlensing to investigate the properties of PBHs (e.g., \cite{esteban2020impact, esteban2023constraints, mediavilla2017limits}).

The angular size of a Type Ia supernova is comparable to that of typical Einstein radii of stars in lensing galaxies. This naturally leads to the idea of using supernovae as an alternative to lensed quasars for dark matter studies with microlensing. Compared to lensed quasars, supernovae present several significant advantages. First, the source size of Type Ia supernovae is more standardized, eliminating the need for complex considerations, such as those required in \cite{awad2023probing}, when using SNe as the source in lensing systems. Second, as `standard candles', the unlensed flux of supernovae at each stage of their explosion is much more predictable. This predictability enables direct use of image flux to study the microlensing effect, rather than relying on flux ratios, as is necessary when quasars are chosen as the source. These advantages enhances the efficiency of statistical analyses\cite{suyu2024strong}. Although the sample size of lensed supernovae has been relatively small in the past, limiting the ability to conduct statistical analysis, the next generation of telescopes,Vera C. Rubin Observatory\cite{suyu2024strong} and Nancy Grace Roman Space Telescope\cite{pierel2021projected}, is highly likely to address this issue, offering a significant increase in the number of detectable events. To be more specific, the Legacy Survey of Space and Time (LSST) from the Rubin Observatory will detect hundreds of SN Ia events per year\cite{oguri2010gravitationally,goldstein2019rates}, which would significantly expand the sample size and enhance the statistical power of future analyses.

In this work, we use microlensing of type Ia supernovae to probe stellar-mass primordial black holes (PBHs). We model PBHs with a point-mass. This is the same as what the literature does, for example, the traditional way to probe PHBs is via  microlensing\cite{mroz2024no,niikura2019microlensing}. By modeling the lens galaxy and its stellar mass distribution, we can calculate the microlensing scatter in the usual manner. Assuming the extreme case where all dark matter consists of PBHs, we generate two scenarios: stars+smooth DM and stars+PBHs. In this context, predictions can be made regarding how many lensed supernova data points are needed to reject one of these cases.

The paper is structured as follows. In Section \ref{2}, we refer to the parameters in a realistic lensed quasar system, PG 1115+80, to provide an initial look at lensed SN Ia and demonstrate how PBHs influence the microlensing scatter. In Section \ref{3}, we simulate the entire lensing process to generate more simulated lensed SN Ia data, allowing us to make the predictions mentioned earlier. Finally, in Section \ref{4}, we will use the mock catalogs from the LSST study \cite{abe2024halo} to give a general calculation and then give discussions and provide our conclusions in Section \ref{5} and \ref{6}.
\begin{figure*}
    \includegraphics[width=2\columnwidth]{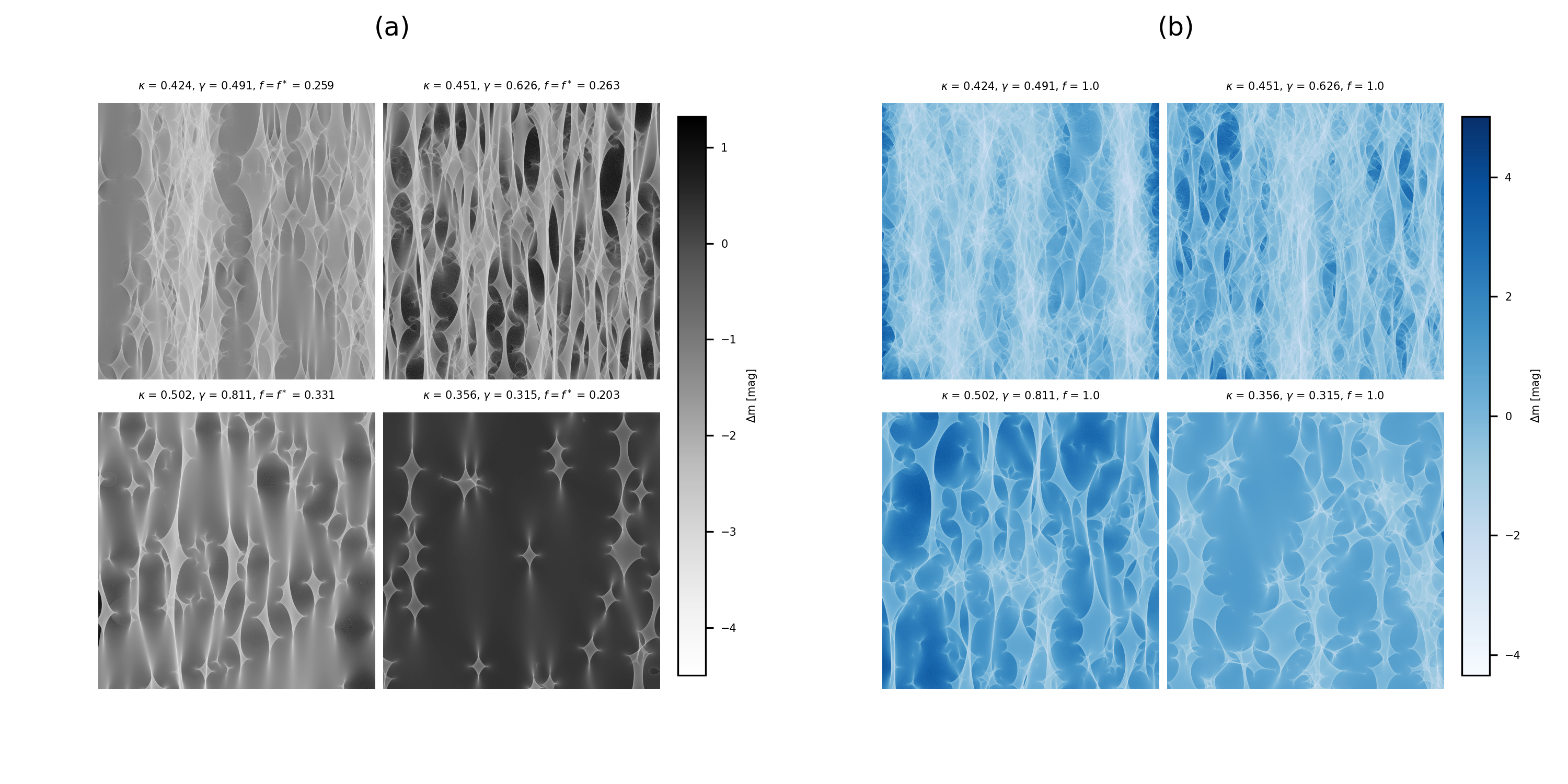}
    \centering
    \caption{Microlensing magnification maps, image (a) represents the stars and smooth DM model, the \, $\kappa,\gamma,f$ comes from the study of PG 1115+080 \cite{chen2019sharp}, while image (b) represents the stars and  PBHs model, we simply set $f \equiv 1$ for that. Both (a) and (b) includes 4 maps corresponding to 4 images of lensed supernova Ia. Every single map gets 4096 pixels each side, equal to $20\mathrm{\langle}R_{\mathrm{Ein}}\mathrm{\rangle}$, per pixel have $1.02 \times 10^{4} \, \mathrm{cm}$ in physical scale. The color gradient, from deep to light, corresponds to the magnitude of the microlensing effect as defined in equation \ref{delta_m}.}
    \label{magmap}
\end{figure*}
\begin{figure*}
    \includegraphics[width=2\columnwidth]{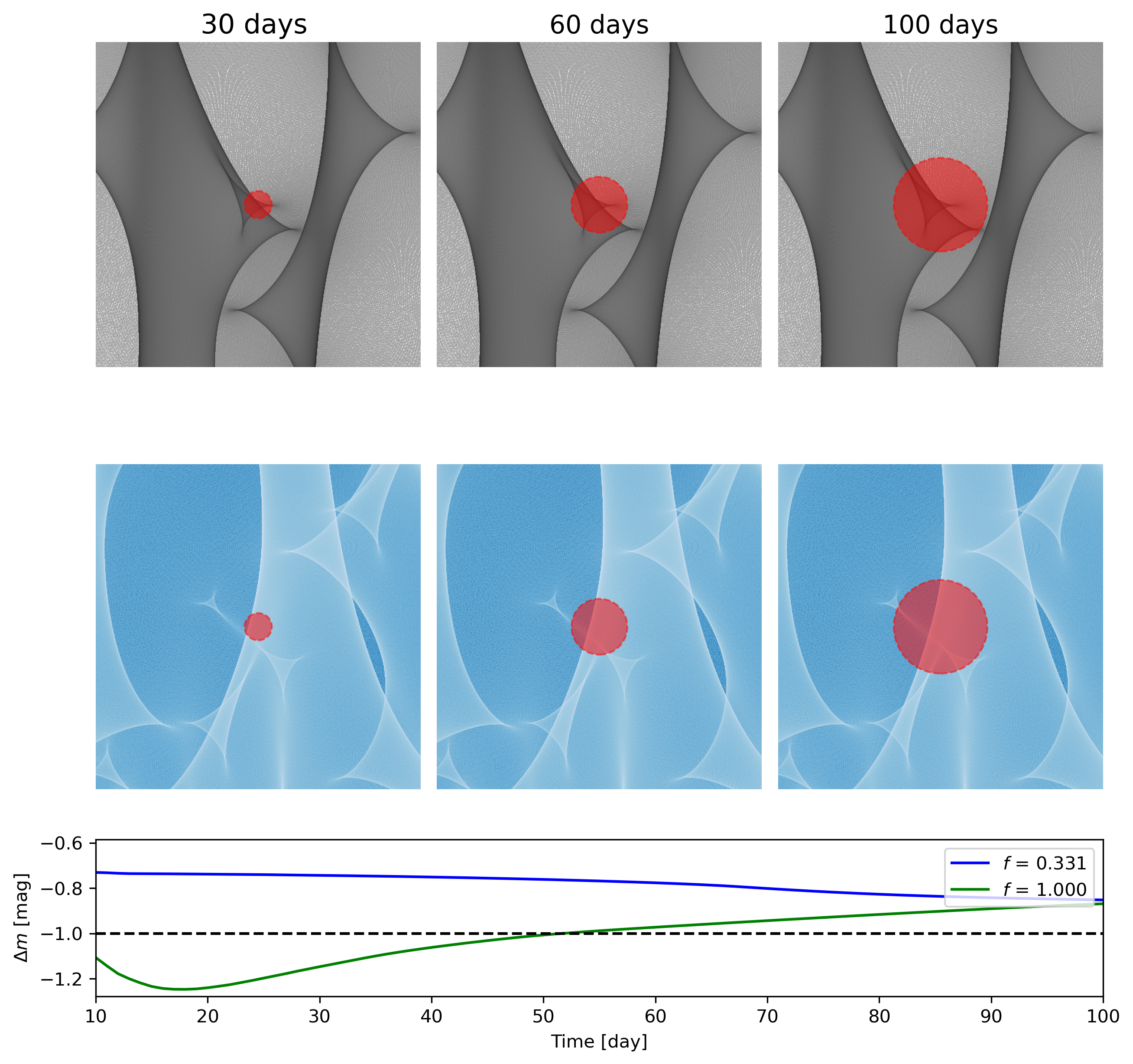}
    \centering
    \caption{The magnification map's zoomed-in view displays an area of \, $1200\times1200$ pixels \, within the magnification map. As the SN Ia profile expands, the magnification values corresponding to the included pixel points within the profile are summed and averaged, therefore, the microlensing perturbation varies over time, as illustrated in the image at the bottom. The gray magnification map and the blue magnification map correspond to different scenarios:stars with smooth DM and stars with PBHs.}
    \label{ML_curve}
\end{figure*}
\begin{figure*}
    \includegraphics[width=2\columnwidth]{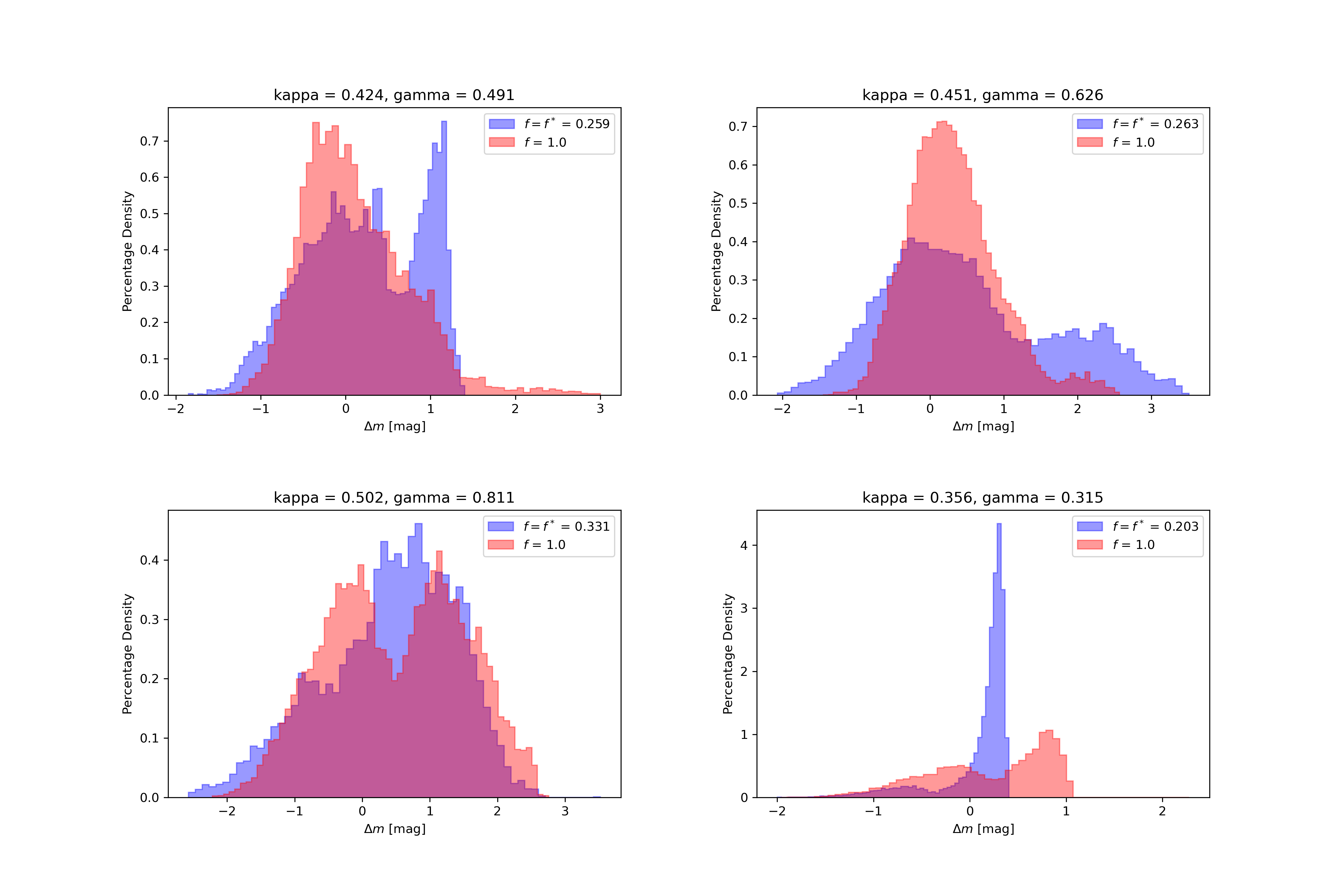}
    \centering
    \caption{Based on our assumptions for PG 1115+080, the histogram of the frequency distribution of the total magnification rates (measured in magnitudes) for the simulated 20000 microlensed supernova events is presented. The definition of the variable on the horizontal axis is provided in \ref{delta_m}, reflecting the deviation of the microlensing magnification from the macro-magnification.}
    \label{hist}
\end{figure*}
\section{The effect of PBHs in microlensing} \label{2}
To probe the stellar-mass PBHs with microlensing, we need to simulate the magnification map. Here we simply consider PBHs as the stellar-mass compact objects modeled with a point-mass model. The simulation of the magnification map is performed using the \textbf{MULES} \footnote{This package is freely available at \href{https://github.com/gdobler/mules}{https://github.com/gdobler/mules.}}  Python package. It is pixelated, with the magnification factor of each pixel calculated using a ray-tracing algorithm\citep{kayser1986astrophysical}. The spatial distribution of stars within the lensing galaxy is random, represented in our magnification map as randomly scattered points.
Three parameters must be generated for microlensing simulation, $\kappa$, $\gamma$, $f$, representing the local environment in which the stars are situated. $\kappa$ is the convergence, which is defined as a dimensionless surface density, while the $\gamma$ is called shear and $f$ is the compact matter fraction. 

According to the review\citep{suyu2024strong}, to date, six lensed supernova events have been successfully observed. These systems
were found either by chance or from targeted programs. None of them are typical lensing systems whose lenses are giant elliptical galaxies for the future survey like LSST. The data for lensed supernovae is so limited that we have to use the lensed quasar system, such as PG 1115+080, and simply replace the source from a quasar to our SN Ia, as this system has been well studied. The local environment for the 4 images had already been given\cite{chen2019sharp}, which is showed in Table \ref{tab:table1}. This section directly uses these parameters.

\begin{table}[b]%The best place to locate the table environment is directly after its first reference in text
\caption{\label{tab:table1}%
Parameters for simulation. $\kappa,\gamma,f$ is given by \cite{chen2019sharp}, we also assume that all dark matter in the lensing galaxy exists in the form of primordial black holes (PBHs), resulting in a compact mass fraction of $f \equiv 1 $ \, in this scenario.
}
\begin{ruledtabular}
\begin{tabular}{lcccc}
\textrm{Map} &
\textrm{Image\,A\,1} & 
\textrm{Image\,A\,2} & 
\textrm{Image\,B} & 
\textrm{Image\,C}  \\
\colrule
$\kappa$ & 0.424  & 0.451  & 0.502 & 0.356  \\
$\gamma$ & 0.491  & 0.626  & 0.811 & 0.315  \\
$f = f^*$ & 0.259   & 0.263 & 0.331 & 0.203 \\
$f = f^*+f_{\mathrm{PBHs}}$ & 1.000   & 1.000 & 1.000 & 1.000\\
\end{tabular}
\end{ruledtabular}
\end{table}

We simulate magnification maps for these four images using both ($\kappa,\gamma,f^*$) and ($\kappa,\gamma,f^*+f_{\mathrm{PBHs}}$), where $*$ stands for the stars. They are showed in Fig.\ref{magmap}. We assume PBHs have the same mass of stars which is set to $\mathrm{\langle}M_{\mathrm{PBHs}}\mathrm{\rangle}=\mathrm{\langle}M_{*}\mathrm{\rangle} = 0.3 M_{\odot}$\cite{liao2015strong}. The maps have a size of $ 20 \mathrm{\langle}R_{\mathrm{Ein}}\mathrm{\rangle} \times 20 \mathrm{\langle}R_{\mathrm{Ein}}\mathrm{\rangle}$ while the pixel resolution is $4096 \times 4096$, $\mathrm{\langle}R_{\mathrm{Ein}}\mathrm{\rangle} \mathrm{=} 2.09 \times 10^{16} \, \mathrm{cm} $ in the source plane, corresponding to redshift configuration (\(z_{\mathrm{s}}=1.72,\,z_{\mathrm{l}}=0.31\)). Here we use the following formula to calculate $\mathrm{\langle}R_{\mathrm{Ein}}\mathrm{\rangle}$:
\begin{equation}
    \mathrm{\langle}R_{\mathrm{Ein}}\mathrm{\rangle} = \sqrt{\frac{4G\mathrm{\langle}M_{*}\mathrm{\rangle}}{c^{2}}\frac{D_{\mathrm{s}}D_{\mathrm{ls}}}{D_{\mathrm{l}}}}.
\end{equation}
Throughout this work, the cosmological model is set to the flat $\Lambda \mathrm{CDM} $ with parameters $H_{0} = 67.8 \, \mathrm{km\,s^{-1} Mpc^{-1}}$, and $\Omega_{\mathrm{m}} = 0.3$\cite{ade2016planck}, respectively. 

This work uses a simple, expanding disk model to simulate the supernova explosion process, thereby obtaining the influence of microlensing on the luminosity of the supernova. According to \cite{mortonson2005size}, the microlensing effect is relatively sensitive to the average size of the source but does not exhibit a strong correlation with specific details in the source's profile model, such as spectral and luminosity distributions. Thus, we place a circular disk with a gradually increasing radius over the magnification map to simulate the supernova. Following Foxley-Marrable et al.\cite{foxley2018impact}, we use an expanding uniform disc to approximate the SN brightness profile. This simple model is sufficient for our purposes since the observed luminosity of a microlensed source is mostly sensitive to the average size and largely independent of any special shape of the source profile\cite{mortonson2005size}.

As an example, we randomly select a point on the magnification map to position the supernova, and we illustrate the variation of the microlensing effect as the disk expands in Fig.\ref{ML_curve}. Since the expansion rate is much larger than the velocities of peculiar motions, we take the microlensing map as time invariant and fix the centroid of the supernova.  
Here we show the microlensing effect using the parameter $\Delta m$,
\begin{equation}
\label{delta_m}
    \Delta m =-2.5\times log_{10}\left|\frac{\mu}{\mu^{\mathrm{SL}}}\right|,
\end{equation}
\begin{equation}
\label{SL}
    \mu^{\mathrm{SL}} = \frac{1}{(1-\kappa)^{2}-\gamma^{2}},
\end{equation}
\(\mu\) is the total magnification, \(\mu^{\mathrm{SL}}\) is the macro magnification, and \(\Delta m\) is the micro magnification, expressed in units of magnitude. 
For SNe Ia, the velocity of expansion does change from the very start of exploration, and before the supernova reach it's peak luminosity about 5 days, the velocity start to decrease linearly, the range of supernova velocity evolution is about $1.5\times10^7\mathrm{m/s}$ to $1.1\times10^7\mathrm{m/s}$. The effect of expansion velocity is embodied in the disk size and as one  will see in Fig.\ref{p_value}, that different sizes of the disk within a certain range have some influence on the final result. A realistic model would give more accurate result. For simply, we set the rate of expansion to $1.5 \times 10^{7} \mathrm{m/s}$ in this work according to \cite{eastman1996atmospheres}. We randomly place 20,000 disks representing the supernova, each corresponding to a size 30 days after the explosion, on the magnification map. This allows us to statistically analyze the distribution of the microlensing effect on the supernova explosion. Every single map have different distribution, they all showed in Fig.\ref{hist}. As we can see, two different models, stars with smooth dark matter and stars with PBHs, do have different distribution in each map. In the following sections of this article, we will present a quantitative study of this difference from a statistical perspective.  
\section{Simulate the lensed SNe Ia }\label{3}
The basic equation in this work is as follows:
\begin{equation}
\label{mag}
\Delta m = -2.5\times log_{10}|\mu^{\mathrm{ML}}|,
\end{equation}
where
\begin{equation}
\label{real}
\mu^{\mathrm{ML}}=\frac{F^{\mathrm{obs}}}{F^{\mathrm{UnL}}\mu^{\mathrm{SL}}},
\end{equation}
\(F^{\mathrm{obs}}\) corresponds to the observed flux result, while \(F^{\mathrm{UnL}}\) represents the unlensed brightness of the type Ia supernova. \(\mu^{\mathrm{SL}}\) denotes the macro magnification. Using equation \ref{mag}, the combination of these observables allows us to calculate the microlensing effect.  

In this work, we aim to determine whether the inclusion of primordial black holes (PBHs) is necessary to account for additional microlensing effects. To do this, we require both \(\Delta m_{\mathrm{sim}}\) and \(\Delta m_{\mathrm{obs}}\). The value of \(\Delta m_{\mathrm{sim}}\) represents the microlensing magnification, in magnitudes, derived from our simulations. We generate a distribution for \(\Delta m_{\mathrm{sim}}\) using the method outlined in section \ref{2}. On the other hand, \(\Delta m_{\mathrm{obs}}\) corresponds to the magnification observed directly. Ideally, we would obtain this value from actual observations, but since such data are not yet available, we instead randomly select values from the \(\Delta m_{\mathrm{sim}}\) distribution to represent \(\Delta m_{\mathrm{obs}}\), while accounting for Gaussian random errors, denoted as \(\sigma_{\mathrm{others}}\) (see subsection B). 

Since \(\Delta m_{\mathrm{sim}}\) is a distribution, whereas \(\Delta m_{\mathrm{obs}}\) is a direct value, they cannot be directly compared. Thus, we require a probability-related quantity to quantify the difference between the two. The method for achieving this will be fully explained in the subsequent sections.

In the remainder of this article, we will consider the following two models:
\paragraph{Fiducial Model}  
We consider the scenario where galaxies are composed of stars and smooth DM to generate both \(\Delta m_{\mathrm{sim}}\) and \(\Delta m_{\mathrm{obs}}\). This model is used to validate whether our method is accurate.  
\paragraph{Alternative Model}  
We consider the scenario where galaxies are composed of stars and smooth DM to generate \(\Delta m_{\mathrm{sim}}\), while \(\Delta m_{\mathrm{obs}}\) is generated assuming galaxies are composed of stars and PBHs. This model represents the central focus of this study.  

This section is structured as follows. First, we provide a detailed description of how the dataset required for this study is generated. Second, we thoroughly describe the method used to statistically distinguish between the two models mentioned above. Finally we will directly show our result in the last subsection.
\subsection{Strong lensing process}\label{3.1}
The upcoming LSST survey is expected to discover numerous lensed SNe Ia events, it's observation strategy had already been studied by \cite{arendse2024detecting}. In this section, we select the mean values from the provided parameter distributions as the parameters for our strong lensing simulation to generate the local environment, \( (\kappa,\gamma,f) \). Here we use the Python package \textbf{Lenstronomy}\cite{birrer2021lenstronomy} \footnote{The official site is \href{https://github.com/lenstronomy}{https://github.com/lenstronomy}} to obtain these three parameters.

We simply set the mass model of the lensing galaxy as a power-law elliptic distribution, with its two-dimensional surface mass density distribution given by:
\begin{equation}
\label{kappa}
    \kappa (x,y) = \frac{3-s}{2}\left(\frac{\theta_{\mathrm{E}}}{\sqrt{q_{\mathrm{lens}}x^2+y^{2}/q_{\mathrm{lens}}}}\right)^{s-1} ,
\end{equation}
while the \(q_{\mathrm{lens}}\) is the axis radio of lens, $ s $ corresponds the profile slope, and \(\theta_{\mathrm{E}}\) denotes the Einstein ring in arcseconds. The origin of the coordinate system is set at the center of the lens, with a rotation angle of \(\phi_{\mathrm{e}}\); external shear is ignored in this work. In this subsection, we set these parameters as follows: $z_{\mathrm{l}}=0.34\,,z_{\mathrm{s}}=0.80\,,q_{\mathrm{lens}} = 0.7 \, , s = 2.0 \, , \theta_{\mathrm{E}} = 0.44''\, , \phi_{\mathrm{e}} = 65^{\circ}\,,\gamma_{\mathrm{ext}}=0\, $, which are all given by \cite{arendse2024detecting}.

\begin{figure}
    \centering
    \begin{minipage}{\linewidth}
		\centering
 		\includegraphics[width=\linewidth]{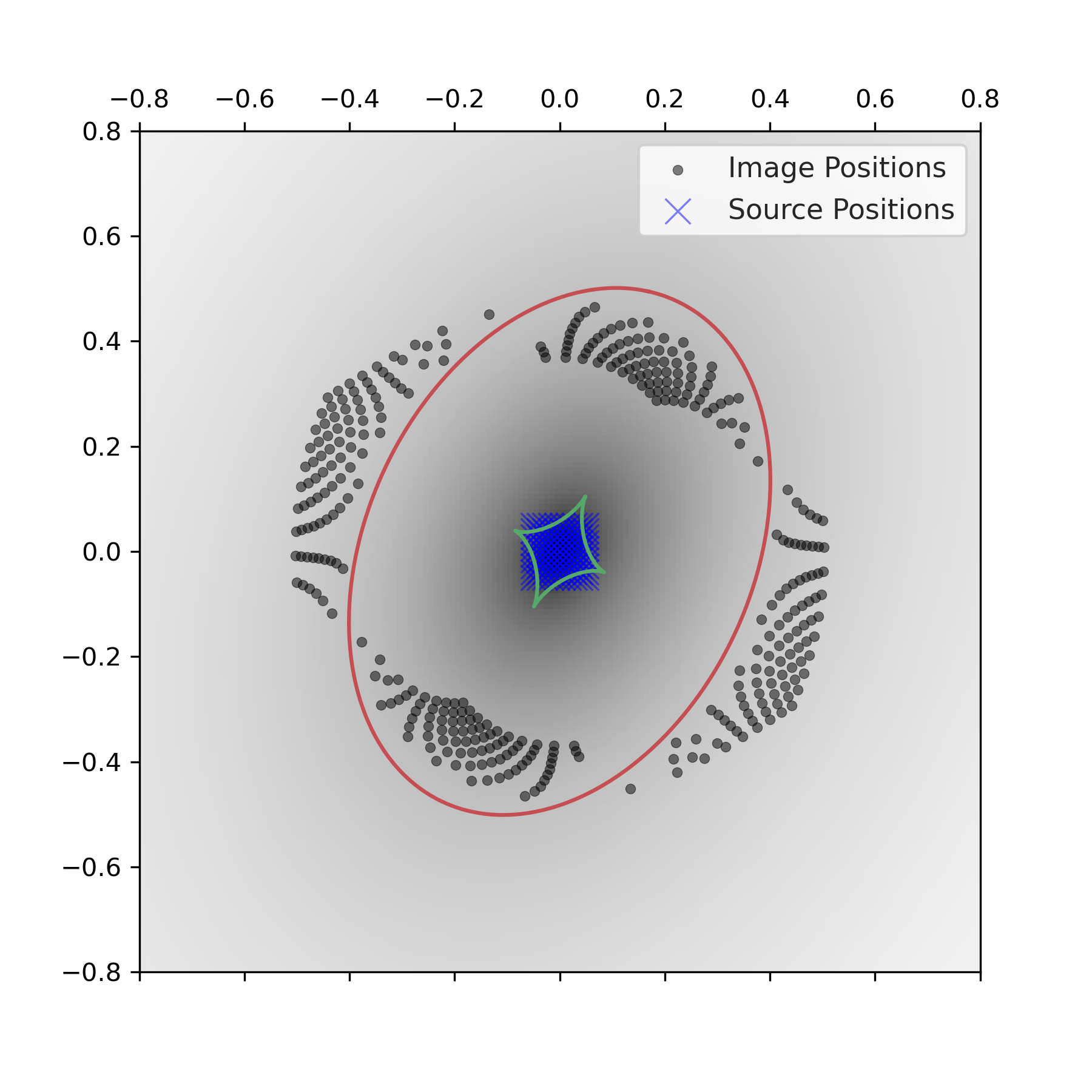}	
 	\end{minipage}
     \caption{A schematic diagram of the strong lensing process. The size of the diagram is \(1.6'' \times 1.6''\), with the background color transitioning from light to dark corresponding to the convergence \(\kappa\), from low to high. The red and green solid lines correspond to the critical line and caustic line of the strong lensing. The blue crosses and black circles represent the positions of the source and its images, respectively.}
     \label{example}
\end{figure}

To generate multiple strongly lensed supernova events, we uniformly select a certain number of positions to place the supernovae (in the strong lensing simulation, the host galaxy is neglected, and the supernova is treated as a point source). We then use the aforementioned Python package to calculate the image positions corresponding to each source location and display them in Fig.\ref{example}.

According to equation \ref{kappa} and the images' positions just calculated, the\,$\kappa$\,and\,$\gamma$\,can \,be rapidly worked out. The next step is to calculate the compact mass fraction, here we assume de Vaucouleurs profile\cite{Dobler_2006}:
\begin{equation}
\label{kappadev}
    \kappa^{*}(x,y) = A e^{-k(r/R_{eff})^{1/4}},  
\end{equation}
which the corresponding complete expression is:
\begin{equation}
\label{fstar}
    f^* = \frac{\kappa^{*}}{\kappa}. 
\end{equation}
Parameters in equation\,\ref{kappadev} are as follows:$k=7.67$\cite{Dobler_2006},$r = \sqrt{q_{\mathrm{lens}}x^2+y^{2}/q_{\mathrm{lens}}}$is the position of the SN image relative to the coordinate origin, \(R_{\mathrm{eff}}\)is effective radius of lens, \(A\) is the normalization factor. The effective radius \(R_{\mathrm{eff}}\) of lens galaxy determined by the following equation,
\begin{equation}
\label{lamda}
    <\Gamma> = \Gamma_{0}+\alpha(lg(\theta_{\mathrm{E}}/R_{\mathrm{eff}})+0.05)+\beta(z_{\mathrm{l}}-0.52).
\end{equation}
According to \cite{li2018strong}, the lens galaxy's mass density gradient is \(\langle \Gamma \rangle \approx 2.00\), and the values of the other parameters are \(\Gamma_{0} = 1.981^{+0.024}_{-0.024}\), \(\alpha = 0.194^{+0.092}_{-0.083}\), and \(\beta = -0.309^{+0.166}_{-0.160}\). For any given position, \(\kappa_{*} \ngeqslant \kappa\)\,and what we want in this work is to identify the differences between the two scenarios mentioned previously, so \(f^*\)\,needs to be minimized as much as possible. Based on these conditions, we can determine value of the normalization factor \(A\), thereby obtaining the expression for \(f^*\).

So far, the preparatory work for simulating \(\mu^{\mathrm{SL}}\) and \(\Delta m_{\mathrm{sim}}\) has been completed. The final subsection of this section will provide a detailed discussion of the final simulation results.
\subsection{Other errors}\label{3.2}
To ensure the highest level of realism in the simulation, a series of necessary errors need to be introduced when simulating the observed supernova magnitudes:
\begin{equation}
\label{errors}
    \sigma_{\mathrm{others}} \sim \sqrt{\sigma^{2}_{\mathrm{SL}}+\sigma^{2}_{\mathrm{int}}+\sigma^{2}_{\mathrm{pho}}}.
\end{equation}

\(\sigma_{\mathrm{SL}}\) represents the strong lensing error, originating from the process of modeling the lens. By assuming a scenario where microlensing can be ignored, we can fit the parameters of the strong lensing process, such as \(\theta_{\mathrm{E}}\), \(q_{\mathrm{lens}}\), and \(s\), as discussed in the previous subsection. This error arises directly from these parameter estimations. In this work, we approximate it as a constant: \(\sigma_{\mathrm{SL}} \approx 0.04 \, \text{mag}\).

\(\sigma_{\mathrm{int}}\) represents the intrinsic error, which arises from the variability in the intrinsic luminosity of SN Ia at a given redshift, \(z_{\mathrm{s}}\). According to \cite{betoule2014improved, scolnic2018complete}, we approximate it as: \(\sigma_{\mathrm{int}} \approx 0.15\, \text{mag}\).

\(\sigma_{\mathrm{pho}}\) represents the photometric error, which comes from the telescope used to detect the SN Ia. For instance, the flux detected by the telescope introduces some errors, and the redshift of SN Ia cannot be measured without uncertainty. Here, we approximate it as: \(\sigma_{\mathrm{pho}} \approx 0.01\, \text{mag}\).

Considering all the above errors, we obtain \(\sigma_{\mathrm{others}} \approx 0.16 \, \text{mag}\), and we will use this value in the next subsection.
\subsection{Method}
After generating the \(\Delta m _{\mathrm{obs}}\) and \(\Delta m _{\mathrm{sim}}\) using the method we have mentioned above, an empirical measurement of the p-value can be used for our statistical analysis. Its form is as follows:
\begin{equation}
\label{pvalue}
    p = 2\times min\{P(\Delta m _{\mathrm{sim}} > \Delta m _{\mathrm{obs}})\,,P(\Delta m _{\mathrm{sim}} < \Delta m _{\mathrm{obs}})\}.
\end{equation}

\begin{figure}
    \centering
    \begin{minipage}{\linewidth}
		\centering
 		\includegraphics[width=\linewidth]{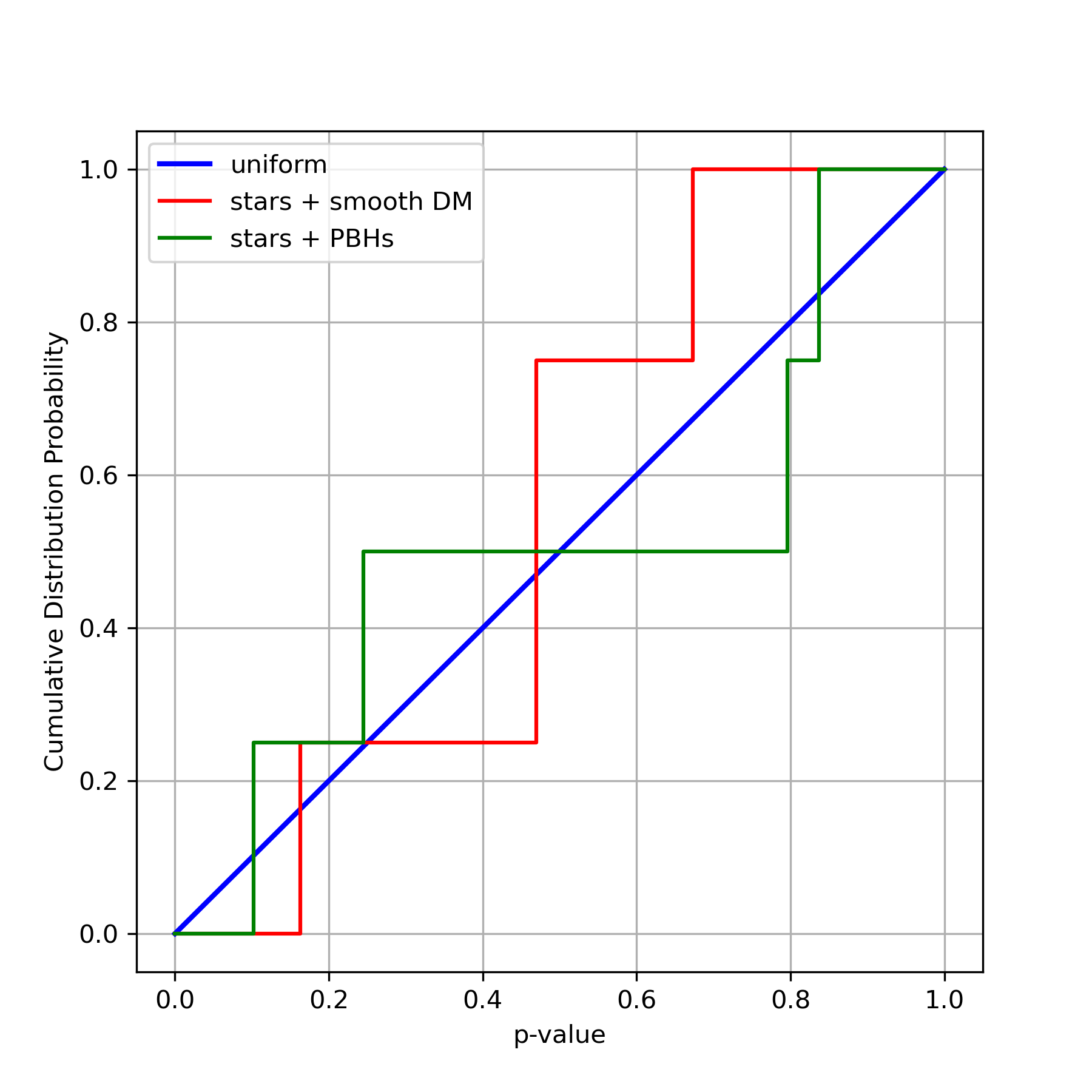}	
 	\end{minipage}
     \caption{The cumulative distribution function of p-values. The red and green solid lines correspond to the two models, stars with smooth DM and stars with PBHs, the blue line is uniform distribution. We performed a KS test using the CDFs of the two models, and the results indicate that both $p_{\mathrm{KS}} > \alpha $. Therefore, based on the current sample size, we cannot reject the hypothesis \(H_0\) in both models.}
     \label{CDF}
\end{figure}

To statistically assess and test the dark matter models, we performed a Kolmogorov-Smirnov (KS) test \cite{chakravarti1967handbook} on the microlensing effects produced by the two models using the p-values defined in equation \ref{pvalue}. When a hypothesis is true, the corresponding p-values are uniformly distributed between 0 and 1\footnote{Proof: \href{https://statproofbook.github.io/P/pval-h0}{https://statproofbook.github.io/P/pval-h0.}}.  

The reason for choosing the KS test over the chi-squared test is primarily that the chi-squared test performs poorly with small sample sizes, whereas the KS test remains sensitive even with limited samples. Here, we define null hypothesis:  
- \(H_0\): The true scenario is stars with smooth DM. But we use two different model, fiducial model-stars with smooth DM and alternative model-stars with PBHs, as \(\Delta m_{\mathrm{obs}}\) to test this hypothesis.

The CDF (Cumulative Distribution Function) used in the KS test is shown in Fig.\ref{CDF}. The KS test compares the separation between the CDFs of the two samples and a uniform distribution to calculate the \(p_{\mathrm{KS}}\). Once \(p_{\mathrm{KS}}\) is obtained, it can be compared with the significance level \(\alpha\). For example, at a 95\% confidence level, if \(p_{\mathrm{KS}} > \alpha = 0.05\), we conclude that we cannot reject the hypothesis at this confidence level.  

As an example, we use the dataset containing four image data points introduced in Section \ref{2} to provide a brief explanation. The results are \(p_{\mathrm{KS}} = 0.472 \gg 0.05\) and \(p_{\mathrm{KS}} = 0.405 \gg 0.05\). This indicates that, based on the four images in our assumption, we cannot statistically reject the null hypothesis. For the fiducial model (stars with PBHs), \(\Delta m_{\mathrm{obs}}\) and \(\Delta m_{\mathrm{sim}}\) come from the same distribution, making \(p_{\mathrm{KS}} > 0.05\) an inevitable result.  

Thus, alternative model is what we need to focus on because determining whether it can be rejected is key to identifying which dark matter model is correct. In Section \ref{3}, we will conduct further simulations to obtain the expected number of supernova events needed to distinguish between the two models.
\subsection{Final results}\label{3.3}
After simulating the strong lensing process using the method described in section \ref{3.1}, we generated numerous parameter pairs, \((\kappa, \, \gamma, \, f^*)\), corresponding to different SN Ia events. We use these parameters to produce various magnification maps and calculate \(\Delta m_{\mathrm{sim}}\) of 30, 60, 100 days after the explosion, following the previously mentioned method, then generate p-values using \ref{pvalue}. Additionally, to make the simulation more realistic, we incorporate the error, \(\sigma_{\mathrm{others}}\), in our KS test.

\begin{figure*}
    \includegraphics[width=2\columnwidth]{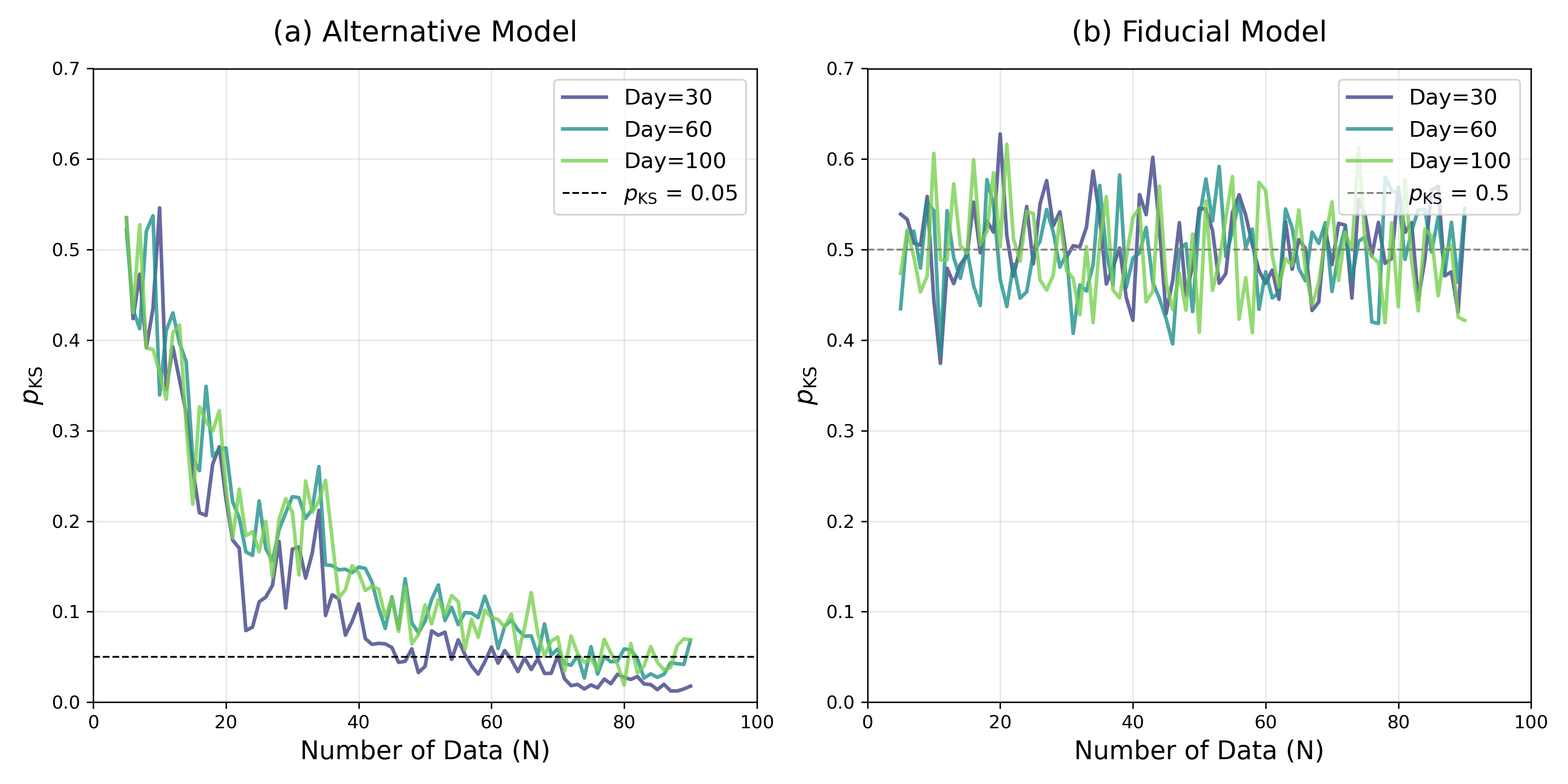}
    \centering
    \caption{The horizontal axis in the plot represents the number of magnification maps used for the KS test, while the vertical axis corresponds to the associated \(p_{\mathrm{KS}}\). For a given number of magnification maps, we perform 100 KS tests and select the median value. Both (a) and (b) show the \(p_{\mathrm{KS}}\) curves for three different stages after the explosion. In (a), which corresponds to the alternative model, we observe that as the number of data points increases to 60, the \(p_{\mathrm{KS}}\) values for all stages drop below 0.05. In (b), the \(p_{\mathrm{KS}}\) values consistently fluctuate around 0.5. These results indicate that the alternative model allows us to reject the null hypothesis after obtaining 60 image data points. However, using the fiducial model, we can never reject the null hypothesis because, in this model, \(\Delta m_{\mathrm{obs}}\) and \(\Delta m_{\mathrm{sim}}\) actually originate from the same distribution.}
    \label{p_value}
\end{figure*}

\begin{figure*}
    \includegraphics[width=2\columnwidth]{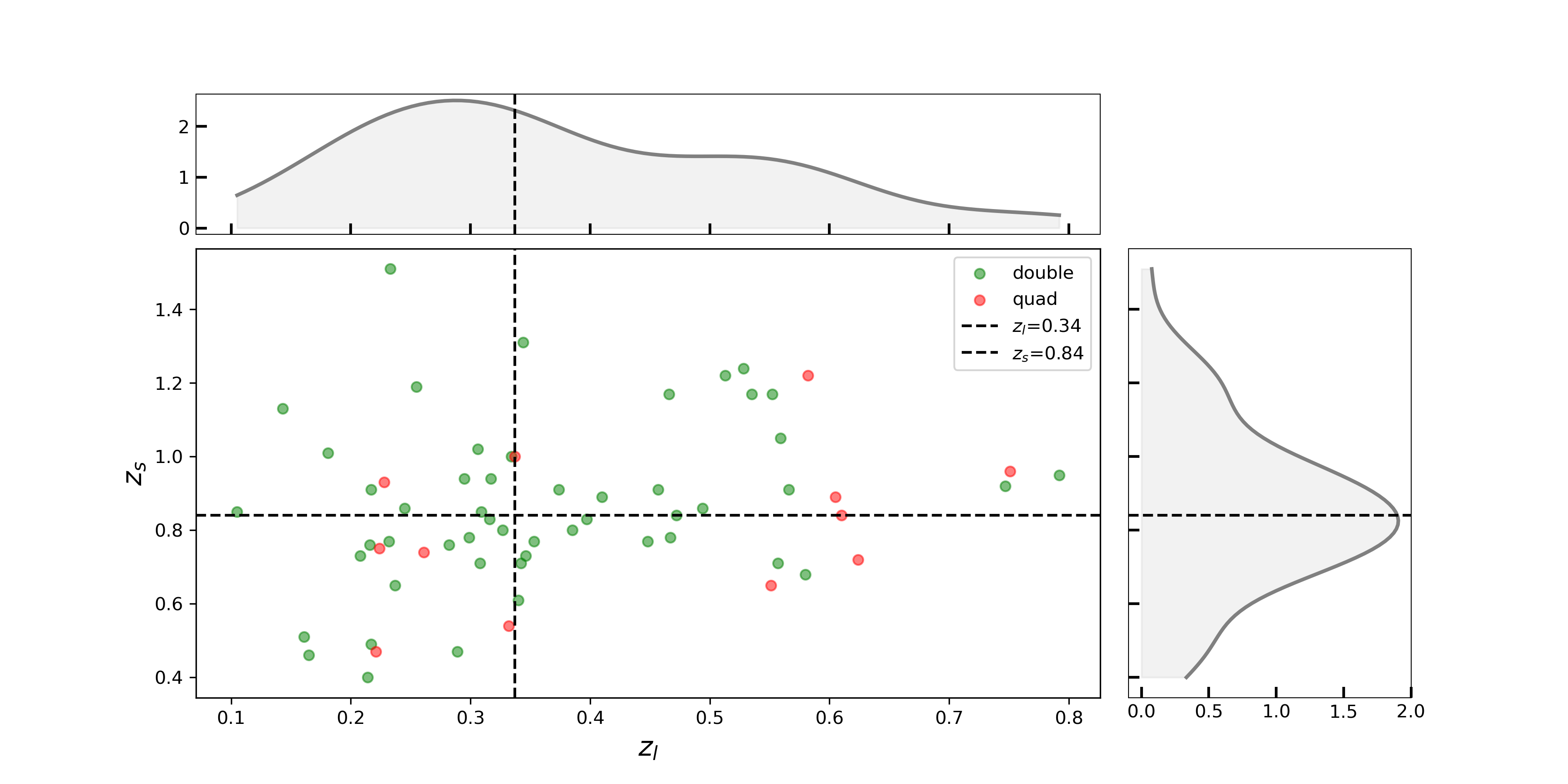}
    \centering
    \caption{The redshift distribution of lensed supernovae Ia is depicted in the plot, with green and red dots representing double and quad systems, respectively. The black dashed lines in the horizontal and vertical directions indicate the median values of the lens and source redshift distributions. Additionally, the upper and right margins of the plot display the fitted density functions for the marginal redshift distributions, providing a clearer view of the distribution trends. }
    \label{redshift}
\end{figure*}

\begin{figure}
    \centering
    \begin{minipage}{\linewidth}
		\centering
 		\includegraphics[width=\linewidth]{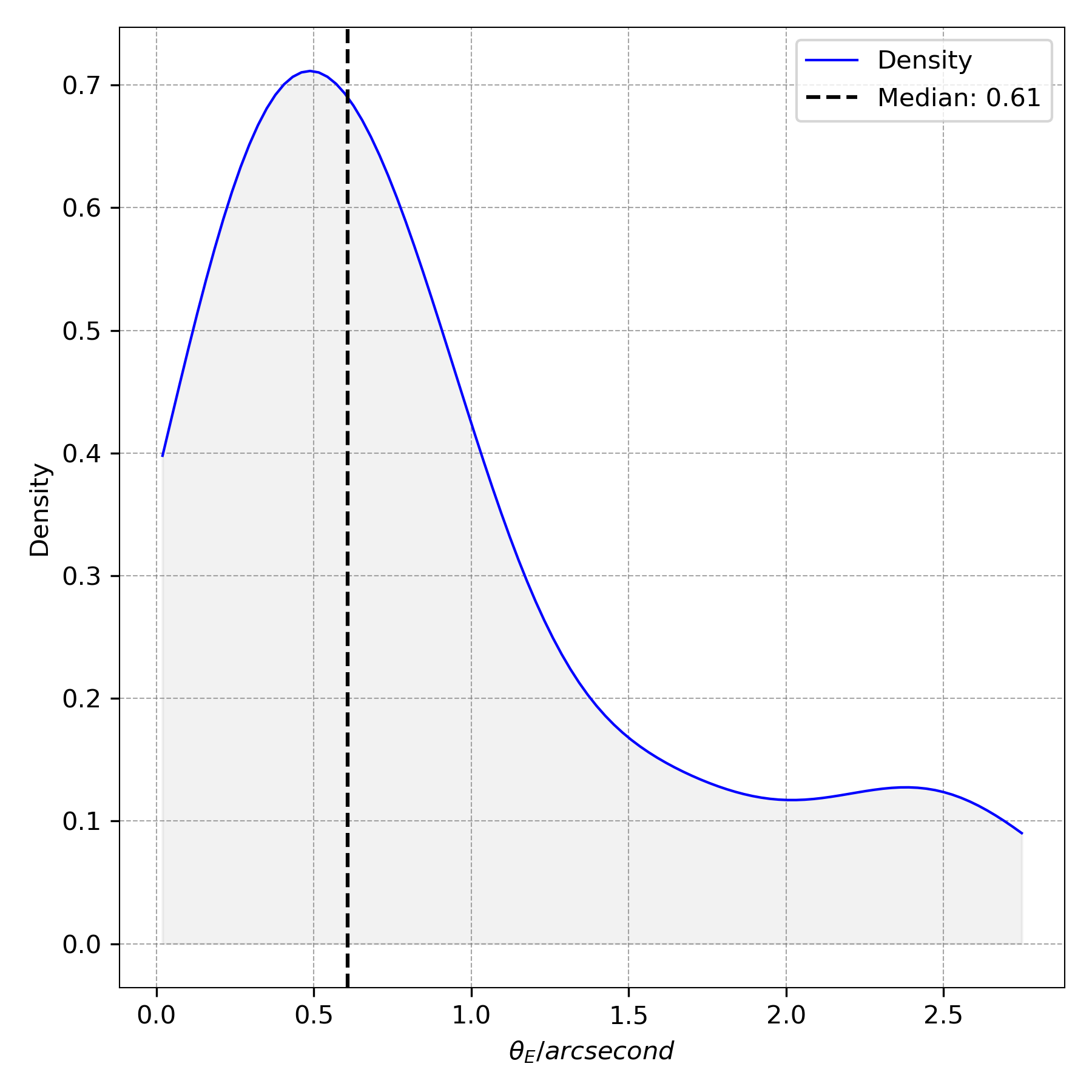}	
 	\end{minipage}
     \caption{The probability density curve of the Einstein radius distribution, fitted using a Gaussian kernel function, is applied to the selected data. The black dashed line represents the median of the Einstein radius distribution.}
     \label{theta}
\end{figure}

\begin{figure*}
    \includegraphics[width=2\columnwidth]{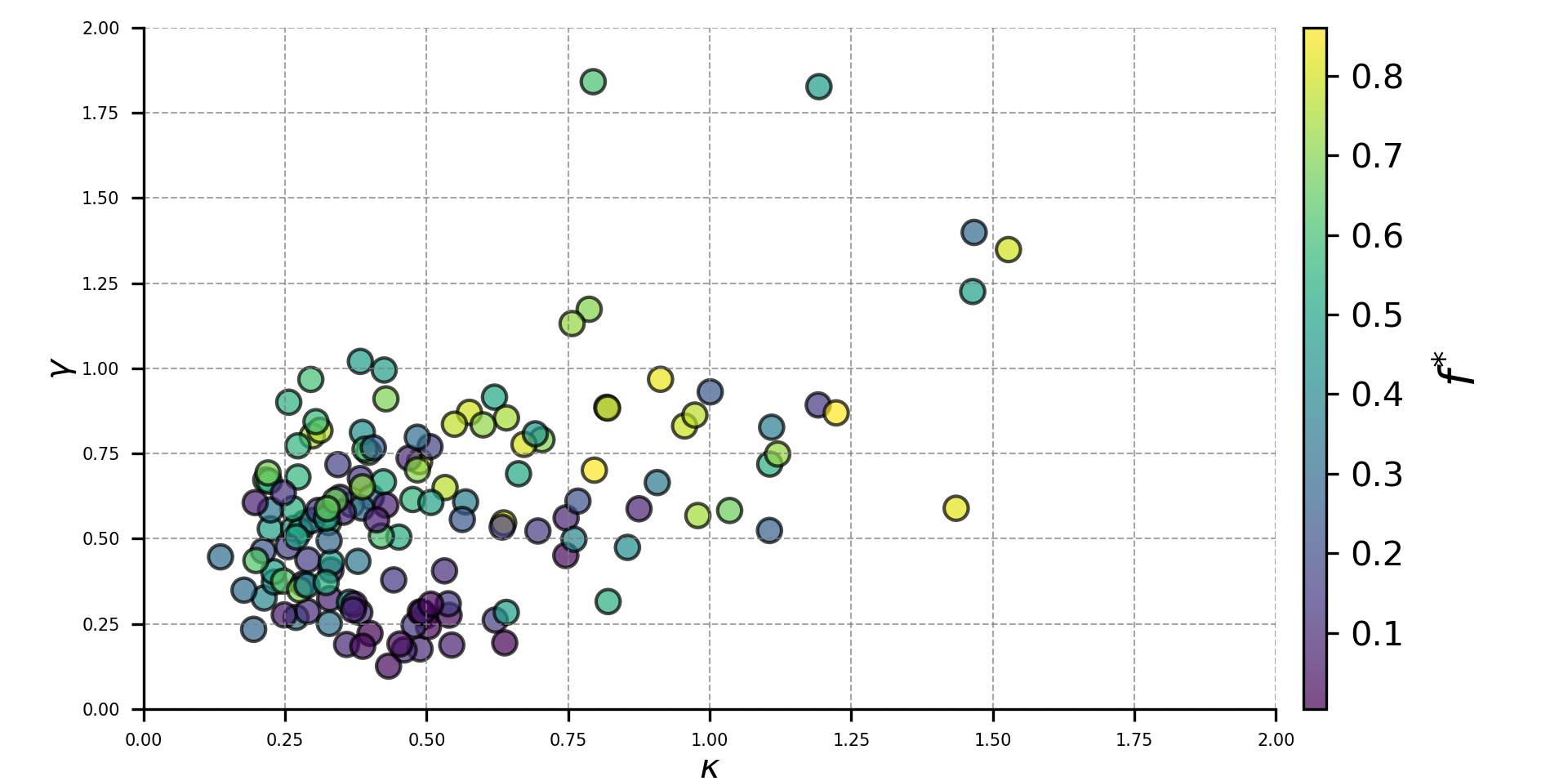}
    \centering
    \caption{The distribution of parameters, \(\left(\kappa,\gamma,f^*\right)\). The horizontal and vertical axes correspond to \(\kappa\) (convergence) and \(\gamma\) (shear). The color of the data points represents the variation in \(f^*\), which indicates the content of compact objects. }
    \label{localenv}
\end{figure*}

\begin{figure*}
    \includegraphics[width=1.5\columnwidth]{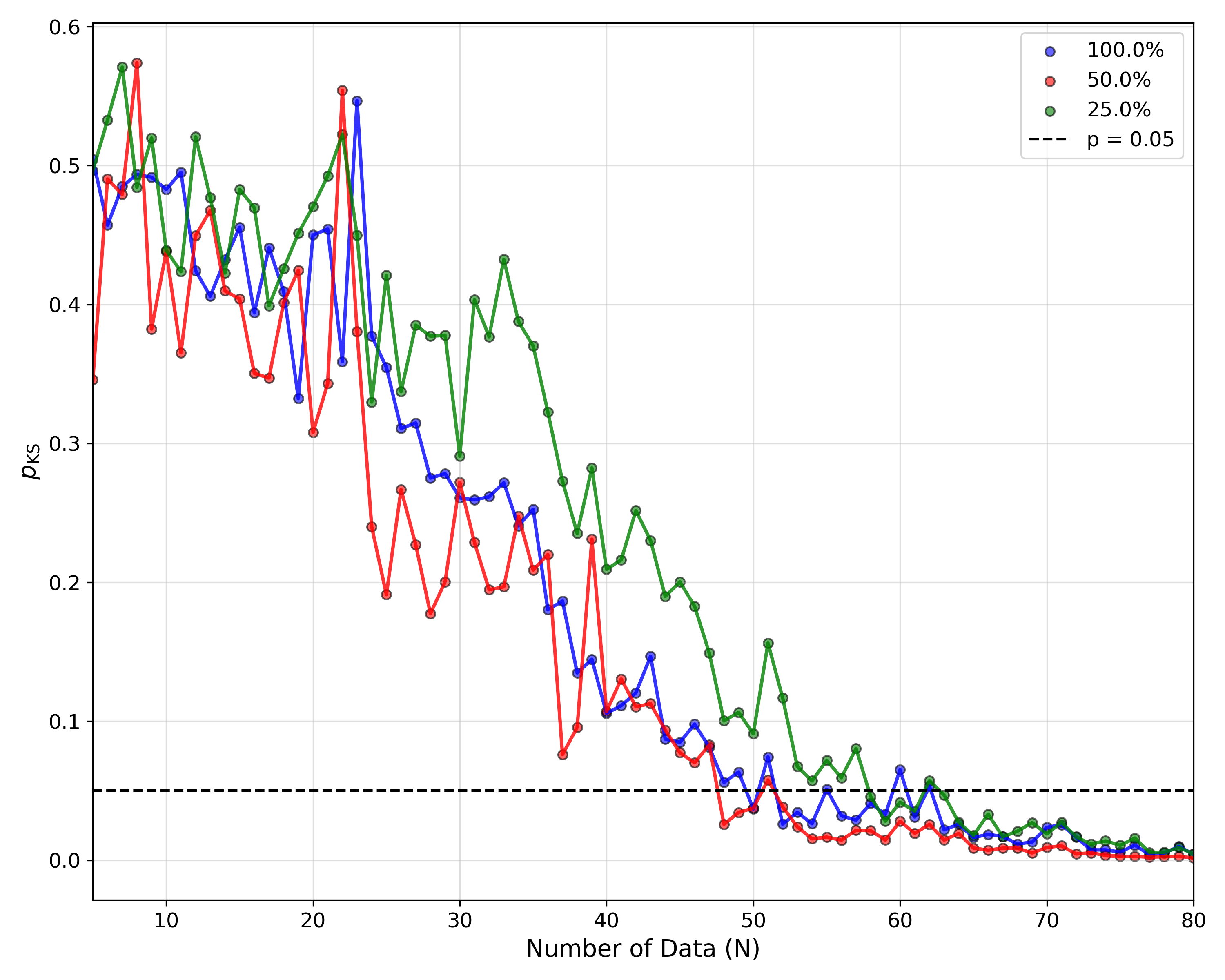}
    \centering
    \caption{The setup of the horizontal and vertical axes is consistent with that of Fig.\ref{p_value}.  The blue, red, and green curves correspond to different compact dark matter fractions. As in Fig.\ref{p_value}, each data point represents the median value obtained after doing KS test 100 times.}
    \label{real_p_value}
\end{figure*}
We plotted the curve of the \(p_{\mathrm{KS}}\) as a function of the number of magnification maps in Fig.\ref{p_value}. For each number of maps, we perform the KS test 100 times and use the median value as our data point. From the plot, it can be observed that for alternative model, although the \(p_{\mathrm{KS}}\) exhibits significant fluctuations initially, it shows an overall decreasing trend. This indicates that as the amount of statistical data increases, the \(p_{\mathrm{KS}}\) generated by the KS test decreases significantly (as expected, since the data used for the p-value in alternative model originates from different models). For the fiducial model, the \(p_{\mathrm{KS}}\) values consistently fluctuate around 0.5, regardless of the number of data points. This indirectly demonstrates the validity of our method.

The plot also reveals that the microlensing effects vary across supernovae at different stages of explosion, consistent with the results presented earlier in Fig.\ref{ML_curve}. Moreover, even with this variability, it can be observed that for \(N > 80\), the \(p_{\mathrm{KS}}\) of each curve for alternative model consistently stabilizes below 0.05. This indicates that, based on the previously mentioned lens galaxy parameters and redshift configurations, it is expected to effectively distinguish between the two dark matter models, stars with smooth DM and stars with PBHs, after obtaining the p-value for 60 image data of lensed supernovae Ia at 30 days after explosion.
\section{Realistic prediction}\label{4}
In the last section, we generated numerous image datasets of type Ia supernovae to provide input for our statistical tool—the KS test—in order to predict how much data is needed to distinguish whether dark matter is composed of smooth objects or 100\% compact objects. However, the method used to produce these image data was not fully rigorous in terms of realistic observations, as each lensed supernova event has its own unique strong lensing parameters. Therefore, in this section, we will use mock catalogs from \cite{abe2024halo}, which have simulated quasars and supernovae events that will be observed by LSST in the future. These catalogs are publicly available on the LSST Strong Lensing Science GitHub \footnote{The official site is \href{https://github.com/LSST-strong-lensing/data\_public}{https://github.com/LSST-strong-lensing/data\_public.}}.

The two main features of these mock catalogs are as follows. First, in contrast to similar mock catalogs like OM10 \cite{oguri2010gravitationally}, these catalogs do not rely on simple mass models such as SIE (Singular Isothermal Ellipsoid) or PL (Power-Law) profiles. Instead, they provide a more detailed mass distribution that combines both dark matter and stellar components. This approach allows the simulations to consider not only galaxy-scale lenses but also group- and cluster-scale lenses. Second, the catalogs take into account the presence of subhalos and satellite galaxies, making the simulations more realistic and reliable.

After analyzing the simulation database, we selected 63 type Ia lensed supernova events from 240 predicted multiple-image lensed supernovae that could be observed by LSST. For each event, we removed the image near the lens center, and the type Ia multiple-image lensed supernova systems were then classified into two categories: double and quad systems.We plot the redshift data and Einstein radii of these systems in the Fig.\ref{redshift} and Fig.\ref{theta}. What's more, the parameters of the local environment are also included in the mock catalogs mentioned above. These are all shown in Fig.\ref{localenv}.

According to section \ref{3.3}, we can distinguish between the two dark matter models using data from 60 images, which correspond to 30 SN Ia events in the case of doubles or 15 SN Ia events in the case of quads. However, as mentioned earlier, there are 63 type Ia lensed supernova events available for analysis, meaning it is more than sufficient in realistic observational scenarios to differentiate between stars with smooth DM and stars with PBHs. In the second model, we assumed that all dark matter consists of compact objects (\(f=f^*+f_{\mathrm{PBHs}} \equiv 1\)), which represents an extremely limiting assumption, as opposed to considering a mixture of smooth DM and PBHs. Given the extensive database provided by the chosen catalogs, more detailed and nuanced analyses can be conducted.  

Here, we assume that the fraction of compact dark matter relative to the total dark matter is \(x\). The compact matter fraction can then be described by the following equation:  
\begin{equation}
f = f^* + x(1-f^*),
\end{equation}
where \(f^*\) corresponds to the scenario of stars with smooth DM.

Due to the limitations of computational power on personal computers, we temporarily refrain from performing a detailed numerical simulation for the precise values of \(x\). Instead, we select three representative values: 1.0, 0.5, and 0.25, corresponding to compact dark matter making up 100\%, 50\%, and 25\% of the total dark matter mass, respectively, for preliminary calculations. Using the method described in Section \ref{3.3}, we can redefine alternative model as : \(\Delta m_{\mathrm{obs}}\) comes from the stars with x fraction of PBHs, incorporating all possible errors as detailed in Section \ref{3.2}. The final results are presented in Fig.\ref{real_p_value}(here we only use alternative model). As we can see in Fig.\ref{p_value}, using different disk sizes for the explosion at various days has little effect on our results when performing the KS test. Therefore, we use the size of 30 days after the explosion for our simulation in Fig.\ref{real_p_value}.

From Fig.\ref{real_p_value}, we can see that as the fraction of compact dark matter (\(x\)) decreases, more image data is required to reject the hypothesis we proposed, with the required numbers being 50, 55, and 65, respectively. When \(x = 1.0\), the required number of image data differs from the 60 data points needed in section \ref{3}. This indicates that the precision of the model has a significant impact on the statistical results.

\section{Discussion}\label{5}
In this section, we will review some of the assumptions made in the simulations presented in this paper, analyze and discuss the results obtained. We will also point out areas for improvement, as well as aspects that need more detailed study in future simulations and observations.

During the past several decades, we have learned more about the class. Observers have discovered a wealth of
explosions that appear to be thermonuclear, but which are outliers from what we normally think of
as the class of SNe Ia. The SNe Ia related to this work are high-redshift ones. In this case, all SNe are almost the normal ones or brighter type (91T-like) which have similar expansion velocities, while 91bg-like SNe Ia who have very low velocities are quite rare. Another type called the high-velocity SNe Ia is rare too at high redshifts, 
see \cite{pan2024measuring} for example.  Thus, in the current theoretical work, we neglect the effect of supernova subtypes.

As mentioned in section \ref{2}, we assume that each part of the supernova explosion is equally affected by the microlensing effect because microlensing is particularly sensitive to the source size rather than the spectral or brightness distribution. Additionally, we have set the average velocity of SNe Ia to \(1.5 \times 10^7 \, \text{m/s}\). This value of the velocity of explosion is much larger than any other velocity like the lens and the source, the velocity of the observer, and the proper motions of the microlenses\cite{foxley2018impact}, for example, the transverse velocity of lensed quasar is \(600\,\mathrm{km/s}\). From Fig. 6, we know different sizes do have some influence on the result. While this assumption would not change our main conclusion, a more realistic disk expansion model would give more accurate result in the future. It is important to note that when using lensed quasars to estimate the stellar-mass PBHs in lens galaxies, as in \cite{awad2023probing}, the uncertainty in the source size has significant adverse effects. This is because the amplitude of microlensing variations decreases as the source size increases \cite{refsdal1991gravitational}. However, the situation is slightly different when using lensed SNe Ia. While the uncertainty in the explosion velocity can also influence the disk size on the magnification map when calculating microlensing scatting, Fig. \ref{p_value} demonstrates that different explosion stages (corresponding to varying disk sizes) have minimal impact on the KS test results. As shown in the figure, all \(p_{\mathrm{KS}}\)-curves for the various stages stabilize below 0.05 once the number of image data reaches 60-70. 

We have considered many potential errors in real observations in section \ref{3}, but some aspects were overlooked. \(\sigma_{\mathrm{SL}}\) is an error strongly correlated with the uncertainty of strong lensing parameters, such as redshifts (\(z_{\mathrm{s}}\,,z_{\mathrm{l}}\)), Einstein radii (\(\theta_{\mathrm{E}}\)), and source positions. Whether we can empirically treat this value as a constant is a topic that warrants further discussion in future studies. Similarly, \(\sigma_{\mathrm{int}}\) and \(\sigma_{\mathrm{pho}}\) require more specific evaluations based on different telescopes and surveys, such as LSST.

Another important issue to note is that, based on the statistical method we use, \(\Delta m_{\mathrm{obs}}\) is randomly selected. Repeating the KS test multiple times to mitigate this randomness is essential. In this study, although we performed 100 tests for each data size (i.e., number of images), as shown in Fig. \ref{real_p_value}, the results still exhibit significant fluctuations. Increasing the number of tests in the future would greatly enhance the precision of the statistical results.

This work is inspired by the study of \cite{awad2023probing}, which used the microlensing of lensed quasars to test whether the stars in the lensing galaxies of a population of lensed quasars are sufficient to account for the observed microlensing signal. The study concluded that, based on the current COSMOGRAIL sample, the two models—stars and smooth DM and stars and PBHs—cannot be distinguished. However, they predict that around 900 microlensing curves would be needed to distinguish between these two models. In contrast, for our work, this number is approximately 25 lensed SNe Ia events.
Thanks to the upcoming LSST survey, approximately 3,500 lensed quasars and 200 lensed supernovae (including one-third being type Ia) with resolved multiple images will be discovered \cite{abe2024halo}. This means that in the future, we will have the opportunity to use both lensed QSOs and SNe Ia for real tests and detailed analyses, allowing us to study the fraction of compact objects relative to the total dark matter mass, using the realistic data from LSST.

In this work, we only use the information of single epoch observations of lensed SNe Ia. In practice, one could get the full light curves of multiple images. We could in principle combine $\Delta m$ from different epochs to get better constraints. The evolution of $\Delta m$ as a function of time (see Fig. 2) contains important information for better constraining PBHs.\\

\section{Conclusion}\label{6}
In this work, we study the microlensing effect caused by stellar-mass compact objects on lensed SNe Ia. More specifically, we provide a statistical prediction for the amount of data required to study the content of stellar-mass PBHs in lensing galaxies. The main results are as follows:

\paragraph{Microlensing Effect}
By using the same image parameters (\(\kappa\), \(\gamma\), \(f\)), we can generally generate microlensing magnification maps. The histograms show that different scenarios exhibit visible differences, but these differences are not significant at a statistical level.
\paragraph{Preliminary Prediction}
By providing certain strong lensing parameters and varying the source position, we can quickly generate a large amount of image data for the KS test. We find that the two models can be distinguished after including 60 image data points, and different explosion stages have minimal influence on this result.
\paragraph{Realistic Prediction}
Using catalogs from the LSST's realistic studies, we assume different compact object fractions relative to the total dark matter mass as 100\%, 50\% and 25\%. We find that approximately 50, 55 and 65 image data points are required to distinguish each of these scenarios, respectively.

Lensed SNe Ia are a powerful tool for studying the nature of dark matter in lensing galaxies. Although the article \cite{mroz2024no} finds no evidence of PBHs in the Milky Way, we cannot rule out the existence of PBHs in other galaxies. The best way to study this problem at cosmological distances is the method outlined in this paper.
\section*{Acknowledgments}
We thank the referee for his/her helpful comments and suggestions. This work was supported by National Natural Science Foundation of China (No. 12222302) and
National Key Research and Development Program of China (No. 2024YFC2207400).
\appendix*

\nocite{*}

\bibliography{Bibs/usedbib}% Produces the bibliography via BibTeX.

\end{document}